\documentclass{article}
\usepackage{amsmath,graphicx,mlspconf}
\usepackage{amssymb,amsfonts,array,multirow,array,hhline,epstopdf,enumerate,bm}
\usepackage{balance}
\allowdisplaybreaks

\def \V{\mathbf{V}}
\def \W{\mathbf{W}}

\def \v{\mathbf{v}}

\def \Rb{{\mathbb R}}
\def \Cb{{\mathbb C}}

\title{deep convolutional neural network-based inverse filtering approach for speech de-reverberation}
\name{Hanwook Chung$^{1,2}$, Vikrant Singh Tomar$^2$ and Benoit Champagne$^1$
\thanks{Funding for this work was provided by grants from NSERC and Mitacs (Canada), with sponsorship from Fluent.ai (Montreal, Canada).}}
\address{
  $^1$ Dept. of Electrical and Computer Engineering, McGill University, Montreal, QC, Canada\\
  $^2$ Fluent.ai, Montreal, QC, Canada\\
email: hanwook.chung@fluent.ai, vikrant.tomar@fluent.ai,  benoit.champagne@mcgill.ca}

\begin{document}
\ninept

%\newcolumntype{D}{>{\centering\arraybackslash}p{2.6em}}
%\newcolumntype{C}{>{\centering\arraybackslash}p{2.2em}}

\newcolumntype{G}{>{\centering\arraybackslash}p{2.2em}} % for smaller font size in tables
\newcolumntype{F}{>{\centering\arraybackslash}p{2.5em}}

\newcolumntype{E}{>{\centering\arraybackslash}p{4.4em}}
\newcolumntype{H}{>{\centering\arraybackslash}p{2.8em}}

\maketitle

\begin{abstract}
In this paper, we introduce a spectral-domain inverse filtering approach for single-channel speech de-reverberation using deep convolutional neural network (CNN). 
The main goal is to better handle realistic reverberant conditions where the room impulse response (RIR) filter is longer than the short-time Fourier transform (STFT) analysis window.
To this end, we consider the convolutive transfer function (CTF) model for the reverberant speech signal.
In the proposed framework, the CNN architecture is trained to directly estimate the inverse filter of the CTF model.
Among various choices for the CNN structure, we consider the U-net which consists of a fully-convolutional auto-encoder network with skip-connections.
Experimental results show that the proposed method provides better de-reverberation performance than the prevalent benchmark algorithms under various reverberation conditions.
\end{abstract}
\begin{keywords}
single-channel speech de-reverberation, inverse filtering, convolutive transfer function, deep convolutional neural network, U-net
\end{keywords}
\section{Introduction}
\label{sec:intro}
When capturing speech from a talker in an enclosed space, a microphone receives multiple delayed and attenuated copies of the original speech signal, caused by the reflections from walls, ceiling and floors, etc. \cite{Naylor10}. 
The general objective of speech de-reverberation algorithms is to remove such reflected components from a reverberant speech signal while preserving the direct-path component to improve its quality and intelligibility.
Speech de-reverberation has been an attractive research area and finds various applications, including mobile telephony, hearing aid and automatic speech recognition. % and source localization.
A considerable amount of research efforts has been devoted to this problem in the past decades, leading to various approaches, such as spectral subtraction \cite{Lebart01, Wu06}, linear prediction-based approahces \cite{Wu06}-\cite{Parchami16} and Kalman filtering \cite{Schwartz14}.
However, these methods were originally introduced by using minimal amount of \emph{a priori} information about the acoustic environment, specified by the room impulse response (RIR) between the speech source and the microphone. %, and the speech signal.
Consequently, they tend to provide limited de-reverberation performance under adverse conditions, e.g., a high level of reverberation or time-varying RIR.

%In recent years, deep neural network (DNN)-based algorithms with strong nonlinear modeling capabilities have attracted enormous interest \cite{Bengio13}.
In recent years, deep learning (DL)-based algorithms with strong nonlinear modeling capabilities have attracted enormous interest \cite{Bengio13}.
They have found diverse applications such as image classification \cite{He16}, automatic speech recognition \cite{Deng13}, speech enhancement \cite{Xu15}-\cite{Chung18} and speech de-reverberation \cite{Han15}-\cite{Qi19}, where they have shown remarkable performance.
%In general, supervised DNN training aims at estimating the nonlinear mapping function that relates the input features to the target features.
In general, supervised DL aims at estimating the nonlinear mapping function that relates the input features to the target features.
In \cite{Han15}, a fully-connected multi-layer perceptron (MLP) is trained to directly predict the clean speech magnitude spectrum from a noisy reverberant speech magnitude spectrum.
This type of approach, which aims to uncover the nonlinear relationship between the input and target features, has been further extended using various deep neural network (DNN) architectures, e.g., long short-term memory (LSTM) units \cite{Zhao18}, convolutional neural network (CNN) and generative adversarial network (GAN) \cite{Ernst18}.
Instead of directly estimating the clean speech features, more robust masking-based approaches have been introduced, e.g., direct estimation of a complex-valued ideal ratio mask (IRM) via MLP \cite{Williamson17}, implicit estimation of a real-valued IRM based on the late reverberation power spectral density (PSD) obatined via MLP \cite{Kodrasi18}, and phase-sensitive mask via GAN \cite{Li18}.
Some references focus on a feature-aware training framework that utilizes additional features as input to the DNN, such as reverberation time \cite{Wu17} or the late reverberation PSD \cite{Qi19}.
The above DNN-based de-reverberation algorithms are implemented in the spectral-domain based on the assumption that the RIR filter length is smaller than the short-time Fourier transform (STFT) analysis window.
In a real world scenario, however, such an assumption is not valid and consequently, provide limited de-reverberation performance.

In this paper, to overcome the above limitation, we introduce a novel spectral-domain inverse filtering approach for single-channel speech de-reverberation using a deep CNN. 
The main goal is to  better handle realistic reverberant conditions, i.e., where the RIR filter length exceeds the STFT window length.
To this end, we consider the convolutive transfer function (CTF) model \cite{Talmon09}, where the reverberant speech spectrum is represented by convolving the clean speech spectral coefficients with spectral filter coefficients along the time frame dimension for each frequency bin.
In the proposed framework, we train the CNN architecture to directly estimate the inverse filter of the CTF model.
During the de-reverberation stage, the estimated inverse filter is applied to the reverberant speech spectrum to obtain the clean speech spectrum.
Among various choices for the CNN structure, we use the U-net consists of a fully-convolutional auto-encoder network with skip-connections \cite{Ronnenberger15}.
Specifically, motivated by \cite{Park17}, we consider an online U-net structure for estimating the inverse filter for each time frame to better handle the time-varying RIR conditions.
Objective experimental results show that the proposed method provides better de-reverberation performance than the prevalent benchmark algorithms under various room reverberation conditions.
%Objective experimental results of source-to-distortion ratio (SDR) \cite{SDR}, extended short-time objective intelligibility \cite{ESTOI} and speech-to-reverberation modulation energy ratio (SRMR) \cite{SRMR} show that the proposed methods provide better de-reverberation performance than the selected benchmark algorithms under various room reverberation conditions.

%\vspace{2em}
\vfill

\section{Reverberant signal model}
\label{sec1}
Let us denote by $y_n$ the observed reverberant speech signal at the discrete-time index $n \in \{0,1,...,N-1\}$.
In the single-channel speech de-reverberation problem, by taking into account the convolutive nature of the acoustic medium as represented by the RIR between the speech source and the microphone, the reverberant speech signal can be written in the time-domain as
\begin{equation}
	y_n = \sum_{q=0}^{Q-1}h_{nq} s_{n-q},
\label{signal_model1}
\end{equation}
where $h_{nq}$ is the time-varying RIR filter coefficient at time $n$ and delay index $q \in \{0,...,Q-1 \}$ and $s_n$ is the clean speech signal.
Considering the propagation delay between the source and microphone, the reverberant speech signal can be divided into three components: direct-path, early reverberant and late reverberant signals.
On this basis, the signal model in (\ref{signal_model1}) can be rearranged as
\begin{equation}
	y_n = \underbrace{\sum_{q=0}^{Q_e-1}h_{nq} s_{n-q}}_{\triangleq y_{n}^{E}} + \underbrace{\sum_{q=Q_e}^{Q-1}h_{nq} s_{n-q}}_{\triangleq y_{n}^{L}},
	\label{signal_model2}
\end{equation}
where $y_n^{E}$ is the sum of the direct-path and early reverberant signals (hereafter referred to as the early reverberant signal for simplicity), $y_n^L$ is the late reverberant signal, and $Q_e$ is the RIR filter length corresponding to early reverberation signal, i.e., the filter index separating the RIR into early and late reverberation components.
%Throughout the paper, we omit the notation ``direct-path'' and refer $y_n^E$ to as early reverberant signal for simplicity.
%Throughout the paper, we simply refer $y_n^E$ to as early reverberant signal.
It has been shown that the late reverberation components are the major cause of the degradation of the speech intelligibility \cite{Nabelek89}. %, e.g., typically selecting $Q_e$ that corresponding to less than 50 ms \cite{Williamson17}.
In this paper, hence, we focus on reducing the late reverberation components, while aiming at recovering the early reverberant signal from the reverberant speech.
Such an algorithm is commonly referred to as late reverberation suppression.
%In this paper, hence, we focus on reducing the late reverberation components, i.e., commonly referred to as late reverberation suppression, while aiming at recovering the sum of the direct-path and early reverberant signals from the a reverberant speech signal.

In audio and speech signal processing, the frequency-domain representation is commonly used in order to better exploit spectral characteristics, where a popular choice is the STFT.
Numerous spectral-domain speech de-reverberation algorithms assume that the RIR filter length is much smaller than the STFT analysis window.
In this case, the effect of reverberation in the frequency domain amounts to a simple multiplication of the room transfer function and the clean speech spectral coefficients.
Such an assumption, however, is often not valid in a real world scenario and hence, may lead to limited de-reverberation performance.
To overcome this limitation, we consider a more comprehensive CTF model \cite{Talmon09}:
\begin{equation}
	Y_{kl} = \sum_{p=0}^{P-1} H_{klp}^{*} S_{k,l-p},
	\label{CTF}
\end{equation}
where $Y_{kl} \in \Cb$ and $S_{kl} \in \Cb$ respectively denote the STFT coefficients of the reverberant and clean speech signals at the frequency bin $k \in \{0,...,K-1 \}$ and time frame $l \in \{0,...,L-1\}$, $H_{klp} \in \Cb$ is the time-varying CTF coefficient with frame delay index $p \in \{0,...,P-1\}$, and the superscript $*$ denotes complex conjugation.
%aim at estimating the inverse filter of the CTF model in (\ref{CTF}), which will be explained in the following section.

\section{Proposed DNN-based  de-reverberation}
In the proposed framework, we aim to directly estimate the inverse filter of the CTF model based on an online U-net architecture.
In this section, after explaining the proposed inverse filtering approach, we describe the online U-net structure and its application to de-reverberation.

\begin{figure*}[t]
\centering
\includegraphics[width = \linewidth]{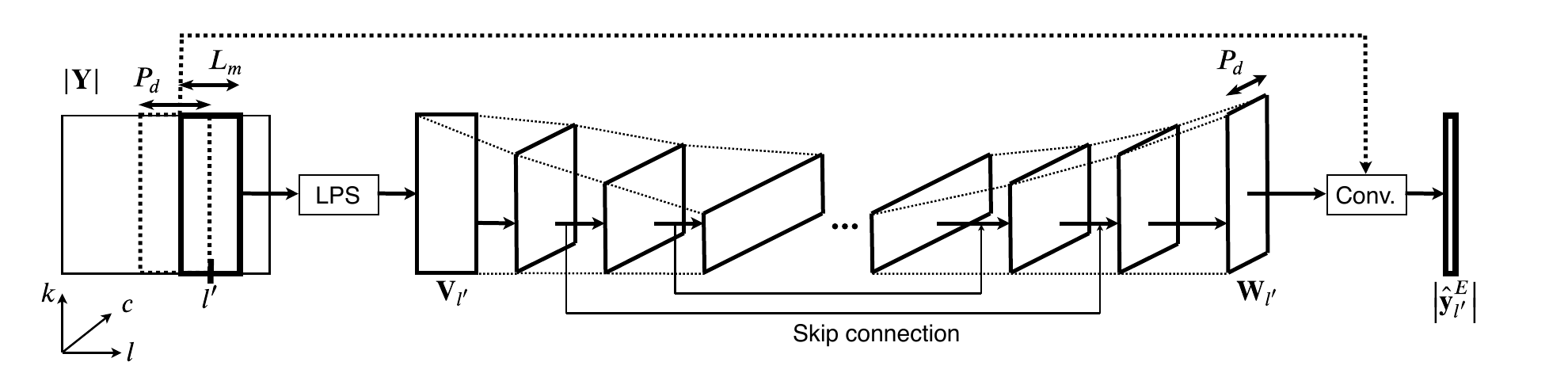}
\caption{\normalsize{The architecture of the online U-net for the proposed inverse filtering-based speech de-reverberation.}}
%\vspace{0.5em}
\label{fig_BD}
\end{figure*}

\subsection{Proposed inverse filtering approach}
\label{sec_ifilt}
The clean speech spectral coefficients, $S_{kl}$, can be estimated via inverse filtering of the CTF model \cite{Li19}:
\begin{equation}
	\hat{S}_{kl} = \sum_{p=0}^{P_d-1} \widetilde{W}_{klp}^{*} Y_{k,l-p},
	\label{ifilt}
\end{equation}
where $\widetilde{W}_{klp} \in \Cb$ is the complex-valued time-varying inverse filter coefficient with frame delay index $p \in \{0,...,P_d-1\}$.
%In this paper, we make two key modifications.
In this paper, we propose a novel inverse filtering method by applying two modifications to (\ref{ifilt}) as follows.
First, instead of estimating the clean speech, we aim at estimating the early reverberant signal, since the latter is sufficient to improve the speech intelligibility as mentioned in Sec. \ref{sec1}.
Furthermore, the suppression of late reverberation is not affected by te misalignment between the observed reverberant and the clean speech signal, which is caused due to the direct-path propagation delay between the source and microphone.
Consequently, this provides a more robust de-reverberation performance.
%\footnote{We indeed verified through experiments that targeting on estimating the early reverberant signal resulted in better de-reverberation performance than targeting on estimating the original clean speech signal.}
Second, instead of estimating complex-valued spectral coefficients, we focus on estimating the spectral magnitudes, as the latter components are knwon to contribute more towards speech intelligibility than the phase components \cite{OShaughnessy87}\footnote{As recent studies taking the phase components into account have shown promising results, e.g., speech enhancement using complex U-net architecture \cite{Choi19}, such an approach remains an interesting avenue for our future work.}.
%Moreover, by using a real-valued inverse filter, we can also reduce the computational cost compared to handling both the magnitude and phase components.
In this way, we can also reduce the computational cost compared to handling both the magnitude and phase components in general.
Hence, taking into account the above modifications, we propose the following inverse filtering model to estimate the magnitude spectral coefficients of the early reverberant signal:
\begin{equation}
	%\left| \hat{Y}_{kl}^{E} \right| = \sum_{p=0}^{P_d-1} W_{klp} \left| Y_{k,l-p} \right|
	|\hat{Y}_{kl}^{E}| = \sum_{p=0}^{P_d-1} W_{klp} |Y_{k,l-p}|,
	\label{prop_ifilt}
\end{equation}
where $W_{klp} \in \Rb$ is the \emph{real-valued} time-varying inverse filter.
%where $W_{klp} \in \Rb$ is the real-valued time-varying inverse filter, which is estimated via an online U-net and will be explained in the following subsection.
%In the proposed framework, the inverse filter is obtained via an online U-net, which will be explained in the following subsection.

\subsection{Online U-net architecture for inverse filtering}
\label{sec_Unet}
%overview:
%1. CNN property and configuration (convolution operations), advantage (local connectivity, shared weights -> less parameters and avoid over-fitting), freq/time/channels
%2. U-net: CAE (fully convolutional layers, bottleneck feature extraction and reconstruction), skip connection (handle vanishing gradient issue and suggests a deeper model)
%3. online U-net: multi-frame input, LPS
%4. output feature

In the proposed framework, we estimate the inverse filter in (\ref{prop_ifilt}) using a CNN structure.
The CNN transforms given input features through a series of hidden layers, based on the convolution operation.
Let $i=\{1,...,I\}$ denote the hidden layer index, while $i=0$ and $i=I+1$ indicate the input and output features, respectively.
The $i$-th hidden layer output is computed by convolving the $(i-1)$-th hidden layer output with a filter, also referred to as kernel, followed by a non-linear transformation via an activation function.
The convolution operation enables to extract local patterns of the given features efficiently, as observed in adjacent time-frequency bins in the STFT domain.
Moreover, the CNN architecture generally requires less parameters, i.e., the number of kernel coefficients, compared to a fully-connected MLP and hence, is known to better handle the over-fitting problem.
The resulting hidden layer output can be represented by a $C_i \times K_i \times L_i$ tensor, where $C_i$, $K_i$ and $L_i$ respectively denote the number of channels, frequency bins and time frames for the $i$-th hidden layer.

Among various choices for the CNN structure, we use the U-net which consists of a fully-convolutional auto-encoder (CAE) with skip-connections \cite{Ronnenberger15}.
%The CAE consists of two stages: \emph{encoder} which compresses the given features into a lower dimensional space, and \emph{decoder} which expands the compressed bottleneck features, that capture the underlying characteristics of the input features, into a desired feature space.
The CAE consists of two stages: an \emph{encoder}, which compresses the given features into a lower dimensional space by capturing their key attributes; and a \emph{decoder}, which expands the compressed features (also known as bottleneck) features into a desired feature space.
The skip-connection method uses the $i$-th hidden layer output as an additional input feature for the $(I-i)$-th hidden layer.
The main advantage of using skip-connection is that it can handle the vanishing gradient issue, which results in an ineffective update of the lower hidden layer parameters due to an extremely small gradient value while implementing error back-propagation.

In the propose framework, motivated by \cite{Park17}, we consider an online U-net structure for estimating an inverse filter for each time to better handle the time-varying RIR conditions.
Regarding the input features, we use the log power spectral coefficients (LPS) of the reverberant speech, i.e., $v_{kl} \triangleq \ln(|Y_{kl}|^2)$.
In order to better handle the temporal dependencies, we construct a multi-frame input feature matrix, $\V_l \in \Rb^{K \times L_m}$, by concatenating feature vectors from $L_m$ successive time frames (e.g., \cite{Park17}) as follows:
\begin{equation}
	\V_l = [\v_{l-(L_m-1)/2},.., \v_l,...,\v_{l+(L_m-1)/2}],
	\label{multV}
\end{equation}
where $\v_l = [v_{kl}] \in \Rb^{K}$ and we consider an odd number for $L_m$.
The output of the online U-net is the inverse filter for the given $l$-th frame, i.e., $\W_l = [W_{klp}] \in \Rb^{K \times P_d}$.
Specifically, we align the frame delay dimension of the inverse filter coefficients along the CNN channel axis for a practical implementation, i.e., $C_{I+1}=P_d$ (a more detailed explanation will be presented in Sec. 4.2).
The online U-net structure for the proposed inverse filtering-based de-reverberation is illustrated in Fig. 1.
%Note that the tensor-form features are expressed in the $k \times l \times c$ coordinate, where $k$, $l$ and $c$ respectively correspond to the frequency bin, time frame and channel.
Note that the tensor form features are expressed in a 3-D $k \times l \times c$ coordinate system, where $k$, $l$ and $c$ respectively indicate the frequency bin, time frame and channel index.

%Note that the online U-net structure enables an efficient real-time implementation while handling time-varying RIR conditions.

During the proposed training stage, the online U-net parameters, i.e., the kernel coefficients, are estimated by minimizing the mean-square error (MSE):
\begin{equation}
	E = \dfrac{1}{KL} \sum_{k=0}^{K-1} \sum_{l=0}^{L-1} \left( |Y_{kl}^{E}|-|\hat{Y}_{kl}^{E}| \right)^2,
\label{MSE}
\end{equation}
where $|\hat{Y}_{kl}^{E}|$ is computed using (\ref{prop_ifilt}). 
During the proposed de-reverberation stage, we estimate the complex-valued early reverberant spectrum by combining the magnitude components estimated via (\ref{prop_ifilt}) and the phase components from the reverberant speech, i.e.,
\begin{equation}
	\hat{Y}_{kl}^{E} = \left( \sum_{p=0}^{P_d-1} \hat{W}_{klp} |Y_{k,l-p}| \right) e^{j\angle Y_{kl}},
\label{de_rev}
\end{equation}
where $j=\sqrt{-1}$.
Finally, the de-reverberated speech signal in the time-domain is reconstructed by applying the inverse STFT to $\hat{Y}_{kl}^{E}$, followed by the overlap-add method.

%During the de-reverberation stage, we first estimate the magnitude spectrum of the early reverberant signal via (\ref{prop_ifilt}).
%Subsequently, the complex-valued early reverberant spectrum is computed by using the phase of the reverberant speech.

\section{Experiments}
In this section, after describing the data sets and general methodology, we present and discuss the experimental results.

\subsection{Data sets}
We conducted experiments using the clean speech from the TIMIT corpus \cite{TIMIT}, where the sampling rate of all signals was 16 kHz.
Regarding the RIR, we employed two data sets: \emph{simulated} RIRs via the RIR generator \cite{RIRHabets}, and \emph{real-measured} RIRs from the C4DM database \cite{RIRC4DM}.
The former are obtained based on the image method for a given 3-D rectangular room, reverberation time $RT_{60}$, and source-microphone positions.
The latter are collected from GreatHall, Octagon and Classroom using the logarithmic sine sweep method, where the measured reverberation time $RT_{30}$ of all rooms was approximately 2s (see \cite{RIRC4DM} for more details).
The speech and RIR files were divided into three \emph{disjoint} groups: i) \emph{training data}, used for estimating the U-net parameters; ii) \emph{validation data}, used for selecting tuning parameters such as the multi-frame length $L_m$ and inverse filter length $P_d$; and iii) \emph{test data}, used during the de-reverberation stage to evaluate the performance. 
For all data, the reverberant speech signals were obtained by convolving the clean speech signals with the RIRs.

Regarding the training data, we selected 4620 utterances from the ``train'' set as the clean speech.
For the simulated RIRs, we considered a room with size of $8 \times 6 \times 4$ m (measured along Cartesian coordinate axes), which will be referred to as Room 1, and reverberation times $RT_{60}$ of 500, 750 and 1000 ms.
We generated 15 RIRs for each reverberation time by varying the source-microphone positions, resulting in a total of 45 RIRs.
For the real-measured RIRs, we selected 50 RIRs from GreatHall, 50 RIRs from Octagon and 40 RIRs Classroom, resulting in a total of 140 RIRs.
Regarding the validation data, we selected 400 utterances from the ``test'' set as the clean speech.
We generated 10 RIRs from Room 1 with $RT_{60}$ of 500, 750 and 1000 ms.
We selected 5 RIRs from each one of GreatHall, Octagon and Classroom.

Regarding the clean speech test data, we selected 192 utterances, from the ``test'' set.
For the simulated RIRs, we generated 10 RIRs from Room 1 with $RT_{60}$ of 500, 750 and 1000 ms.
To evaluate the performance for an unseen type of acoustic environment, we additionally generated 10 RIRs from a room with size of $6 \times 4 \times 3.5$ m, which will be referred to as Room 2, and reverberation times $RT_{60}$ of 500, 750 and 1000 ms.
For the real-measured RIRs, we selected 10 RIRs from each one of GreatHall, Octagon and Classroom.
Besides the above \emph{static} RIR conditions, we additionally considered \emph{time-varying} RIR scenarios.
To this end, we divided 10 RIRs into two groups, allowing us to generate two different time-varying RIRs scenarios, each comprised of 5 RIRs.
Specifically, for each scenario, the reverberant speech was obtained by convolving the given speech utterance with one of the 5 RIRs in cycle for every 1s.

\setlength\extrarowheight{2pt}
\begin{table}[t!]
%\footnotesize
\scriptsize
%\fontsize{7}{8}\selectfont % same as scriptsize
%\fontsize{6}{8}\selectfont
%\fontsize{6.5}{7.5}\selectfont
\caption{Average results for the static simulated RIRs}
\centering
	%\begin{tabular}{c|c|c|c|c|c|c||c}
	\begin{tabular}{G|G|F|F|F|F|F||F}
		\hline
%		Room & \multirow{2}{*}{$RT_{60}$} & \multirow{2}{*}{Eval.} & \multirow{2}{*}{Rev.} & \multirow{2}{*}{DSM} & \multirow{2}{*}{iIRM} & \multirow{2}{*}{dIRM} & \multirow{2}{*}{iFilt} \\ %\cline{7-8}
		Room & $RT_{60}$ & \multirow{2}{*}{Eval.} & \multirow{2}{*}{Rev.} & \multirow{2}{*}{DSM} & \multirow{2}{*}{iIRM} & \multirow{2}{*}{dIRM} & \multirow{2}{*}{iFilt} \\ %\cline{7-8}
		type & (ms) & & & & & & \\
		\hline \hline
		\multirow{9}{*}{\rotatebox[origin=c]{90}{Room 1}} &
		\multirow{3}{*}{500}
            	& SDR & 2.73 & 2.73 & 3.86 & 1.35 & \textbf{4.19} \\ %\cline{2-10}
            	& & ESTOI & 0.49 &0.62 & 0.52 &	0.58 & \textbf{0.67} \\ %\cline{2-10}
            	& & SRMR & 2.69 &2.98 &2.63 &	2.41 & \textbf{3.59} \\ \cline{2-8}
		%\hline %\hline
		& \multirow{3}{*}{750}
            	& SDR & -0.56 & 1.11 & 	1.27 & 0.14 & \textbf{2.25} \\ %\cline{2-10}
            	& & ESTOI & 0.33 & 0.52 & 0.38 & 0.50 & \textbf{0.55} \\ %\cline{2-10}
            	& & SRMR & 2.14 & 2.61 & 2.23	 & 2.28 & \textbf{3.24} \\ \cline{2-8}
		%\hline %\hline
		& \multirow{3}{*}{1000}
            	& SDR & -2.52 & -0.20 & -0.59 & -0.52 & \textbf{0.93} \\ %\cline{2-10}
            	& & ESTOI & 0.26 & 0.46 & 0.30 & 0.45 & \textbf{0.48} \\ %\cline{2-10}
            	& & SRMR & 1.79 & 2.39 & 1.97 & 2.24 & \textbf{3.01} \\ %\cline{2-10}
		\hline \hline
		
		\multirow{9}{*}{\rotatebox[origin=c]{90}{Room 2}} &
		\multirow{3}{*}{500}
            	& SDR & 3.17 & 2.81 & 4.02 & 1.07 & \textbf{4.19} \\ %\cline{2-10}
            	& & ESTOI & 0.49 & 0.63 & 0.53 & 0.57 & \textbf{0.67} \\ %\cline{2-10}
            	& & SRMR & 2.77 & 3.03 & 2.72	 & 2.41 & \textbf{3.66} \\ \cline{2-8}
		%\hline %\hline
		& \multirow{3}{*}{750}
            	& SDR & 0.58 & 1.83 & 2.45 & 0.52 & \textbf{3.12} \\ %\cline{2-10}
            	& & ESTOI & 0.36 & 0.55 & 0.41 & 0.53 & \textbf{0.59} \\ %\cline{2-10}
            	& & SRMR & 2.21 & 2.81 & 2.40 & 2.36 & \textbf{3.45} \\ \cline{2-8}
		%\hline %\hline
		& \multirow{3}{*}{1000}
            	& SDR & -1.64 & 0.43 & 0.48 & -0.27 & \textbf{1.58} \\ %\cline{2-10}
            	& & ESTOI & 0.28 & 0.49 & 0.32 & 0.48 & \textbf{0.51} \\ %\cline{2-10}
            	& & SRMR & 1.83 & 2.56 & 2.08 & 2.29 & \textbf{3.18} \\ %\cline{2-10}
		\hline
	\end{tabular}
	\label{Habets_static_RIR}
\end{table}

\subsection{Methodology}
Regarding the implementation of the proposed de-reverberation algorithm, a Hamming window of 400 samples with 60\% overlap and 512-point fast Fourier transform (FFT) were employed for the STFT analysis.
We set the multi-frame length to $L_m=5$, the inverse filter length to $P_d=9$, and the early reverberation filter lengths to $Q_e \in \{32, 64, 128\}$.
We designed the U-net using $I=11$ hidden layers with corresponding numbers of channels $[16,16,32,32,64,64,64,32,32,16,16]$ and $C_{I+1}=P_d=9$ for the output layer, based on the validation data.
We used the rectified linear units (ReLU) as the activation function for all hidden layers and linear activation function for the output layer.
Batch normalization was applied to all hidden layers \cite{Loffe15}.
The U-net parameters were updated iteratively via error back-propagation and the adaptive moment estimation (Adam) optimizer \cite{Kingma14}, with the mini-batch size of 32 for a total of 200 epochs.
The initial learning rate was set to 0.001, which decreased by 10\% for every 10 epochs.
We used the kernel size of 9 with stride of 2 along the $k$-axis for all hidden and output layers.
To ensure the non-negativity of the estimated spectral magnitudes of the early reverberant speech, i.e., $|\hat{Y}_{kl}^{E}| \ge 0$, we applied ReLU to the inverse-filtered reverberant speech in (\ref{prop_ifilt}). 
In order to efficiently implement the proposed online U-net for inverse filtering, we considered the 2-D convolution operation for all layers as follows.
For the given input feature matrix $\V \in \Rb^{K \times L}$, we applied the kernels of sizes $9 \times L_m$ to the input layer $i=0$, and $9 \times 1$ to all hidden layers $i \in \{1,...,I\}$, with stride of 2 along the $k$-axis and 1 along the $l$-axis.
%For a given input feature $\V \in \Rb^{K \times L}$, we applied the kernel of size $9 \times L_m$ with stride of $(2,1)$ for $i=1$ and $9 \times 1$ with the stride of $(2,1)$ for $i=\{2,...,I+1\}$, where the dimensions of the strides are expressed in the $k-l$ axis.
The number of time frames was set to $L_i=L$ for all $i \in \{ 0,1,...,I \}$, and the size of the output layer was $C_{I+1} \times K_{I+1} \times L_{I+1}=P_d \times K \times L$.
The time-shifted magnitude spectra of the reverberant speech were concatenated along the $c$-axis to obtain a tensor of size $C_{0} \times K_{0} \times L_{0}=P_d \times K \times L$.
The estimated early reverberant magnitude spectrum given by (\ref{prop_ifilt}) was then computed by multiplying the above U-net input and output tensors, followed by summing along the $c$-axis.

\setlength\extrarowheight{2pt}
\begin{table}[t!]
%\footnotesize
\scriptsize
%\fontsize{7}{8}\selectfont % same as scriptsize
%\fontsize{6}{8}\selectfont
%\fontsize{6.5}{7.5}\selectfont
\caption{Average results for the time-varying simulated RIRs}
\centering
	%\begin{tabular}{c|c|c|c|c|c|c||c}
	\begin{tabular}{G|G|F|F|F|F|F||F}
		\hline
%		Room & \multirow{2}{*}{$RT_{60}$} & \multirow{2}{*}{Eval.} & \multirow{2}{*}{Rev.} & \multirow{2}{*}{DSM} & \multirow{2}{*}{iIRM} & \multirow{2}{*}{dIRM} & \multirow{2}{*}{iFilt} \\ %\cline{7-8}
		Room & $RT_{60}$ & \multirow{2}{*}{Eval.} & \multirow{2}{*}{Rev.} & \multirow{2}{*}{DSM} & \multirow{2}{*}{iIRM} & \multirow{2}{*}{dIRM} & \multirow{2}{*}{iFilt} \\ %\cline{7-8}
		type & (ms) & & & & & & \\
		\hline \hline
		\multirow{9}{*}{\rotatebox[origin=c]{90}{Room 1}} &
		\multirow{3}{*}{500}
            	& SDR & -0.84 & 0.00 & 0.42 & -0.34 & \textbf{0.73} \\ %\cline{2-10}
            	& & ESTOI & 0.48 & 0.62 & 0.52 & 0.58 & \textbf{0.66} \\ %\cline{2-10}
            	& & SRMR & 2.71 & 2.96 & 2.67 & 2.48 & \textbf{3.61} \\ \cline{2-8}
		%\hline %\hline
		& \multirow{3}{*}{750}
            	& SDR & -3.30 & -1.01 & -1.41 & -1.04 & \textbf{-0.12} \\ %\cline{2-10}
            	& & ESTOI & 0.34 & 0.52 & 0.38 & 0.50 & \textbf{0.55} \\ %\cline{2-10}
            	& & SRMR & 2.10 & 2.61 & 2.18 & 2.31 & \textbf{3.23} \\ \cline{2-8}
		%\hline %\hline
		& \multirow{3}{*}{1000}
            	& SDR & -5.09 & -2.72 & -3.28 & -2.62 & \textbf{-1.90} \\ %\cline{2-10}
            	& & ESTOI & 0.24 & 0.45 & 0.28 & 0.45 & \textbf{0.47} \\ %\cline{2-10}
            	& & SRMR & 1.74 & 2.37 & 1.94 & 2.21 & \textbf{2.97} \\ %\cline{2-10}
		\hline \hline
		
		\multirow{9}{*}{\rotatebox[origin=c]{90}{Room 2}} &
		\multirow{3}{*}{500}
            	& SDR & -0.61 & -0.46 & \textbf{0.16} & -1.71 & 0.12 \\ %\cline{2-10}
            	& & ESTOI & 0.49 & 0.63 & 0.53 & 0.57 & \textbf{0.67} \\ %\cline{2-10}
            	& & SRMR & 2.76 & 2.97 & 2.69 & 2.37 & \textbf{3.58} \\ \cline{2-8}
		%\hline %\hline
		& \multirow{3}{*}{750}
            	& SDR & -2.92 & -1.57 & -1.30 & -2.39 & \textbf{-0.90} \\ %\cline{2-10}
            	& & ESTOI & 0.32 & 0.53 & 0.38 & 0.51 & \textbf{ 0.56} \\ %\cline{2-10}
            	& & SRMR & 2.11 & 2.77 & 2.28 & 2.33 & \textbf{3.35} \\ \cline{2-8}
		%\hline %\hline
		& \multirow{3}{*}{1000}
            	& SDR & -3.96 & -1.79 & -1.94 & -2.07 & \textbf{-1.01} \\ %\cline{2-10}
            	& & ESTOI & 0.26 & 0.47 & 0.31 & 0.46 & \textbf{0.49} \\ %\cline{2-10}
            	& & SRMR & 1.72 & 2.43 & 1.98 & 2.21 & \textbf{3.03} \\ %\cline{2-10}
		\hline
	\end{tabular}
	\label{Habets_tv_RIR}
\end{table}

To evaluate the performance of the proposed method, we implemented several benchmark algorithms: i) direct estimation of the clean speech magnitude spectral coefficients (DSM) \cite{Ernst18}, ii) implicit estimation of a real-valued IRM based on the late reverberation PSD obtained via DNN (iIRM) \cite{Kodrasi18}, and iii) direct estimation of a real-valued IRM (dIRM) \cite{Wang14}.
Specifically, we considered the LPS of the early reverberant signal as the target output feature for DSM.
The target IRM in the dIRM method was constructed based on the well-known Wiener filter, specified by the PSDs of the early and late reverberant signals, i.e., $|Y_{kl}^{E}|^2/(|Y_{kl}^{E}|^2 + |Y_{kl}^{L}|^2)$.
Although the iIRM and dIRM methods were originally proposed using a fully-connected MLP, we implemented them using the online U-net as explained in Sec. 3.2 for fair comparison.
Basic settings such as the STFT analysis and synthesis, the U-net configuration, the input feature type (i.e., the LPS of the reverberant speech) and the mini-batch size were kept identical when applicable.

To evaluate the de-reverberation performance, we considered the source-to-distortion ratio (SDR) \cite{SDR}, extended short-time objective intelligibility (ESTOI) \cite{ESTOI} and speech-to-reverberation modulation energy ratio (SRMR) \cite{SRMR} as the objective measures.
The SDR is computed in dB based on the source-to-interference ratio (SIR) and source-to-artifact ratio (SAR), and has been widely used in audio source separation and speech enhancement, e.g., \cite{Park17, Chung18}.
For a given target source signal, in general, the interference refers to unwanted signal such as late reverberation components, whereas the artifact refers to forbidden distortion.
In speech de-reverberation applications, these measures can be interpreted as follows: the SIR and SAR are proportional to the amount of late reverberation suppression and inversely proportional to the clean speech distortion, respectively, while the SDR measure the overall quality of the de-reverberated speech signal.
The ESTOI is computed based on the spectral correlation between the short-time auditory filter-bank coefficients of the target clean speech and processed speech, and has shown to be closely related to speech intelligibility of a human listener.
The SRMR is a non-intrusive metric for speech quality and intelligibility based on an auditory-inspired modulation spectral representation of the speech signal. 
%variability reduction and improved intelligibility estimation both for normal hearing listeners and cochlear implant users.
For all measures, a higher value indicates a better result.

\subsection{Results}
The average results using the static and time-varying simulated RIR, in case of an early reverberation RIR filter length of $Q_e=32$, are shown in Tables \ref{Habets_static_RIR} and \ref{Habets_tv_RIR}, respectively.
The proposed inverse filtering-based approach is referred to as iFilt.
The values in bold indicate the best performance along the corresponding row.
As we can see, the propose method provided better de-reverberation performance than the benchmark algorithms for all room types and both the static and time-varying RIR conditions, in general.
The only exception was found from the SDR value for the time-varying RIR in Room 2 at $RT_{60}=500$ ms, where the iIRM method provided slightly better result.

%As we can see, the proposed method provided better de-reverberation performance than the benchmark algorithms for all room types and both the static and time-varying RIR conditions, in general.
%The only exceptional case was found from the SDR value for the time-varying RIR in Room 2 at $RT_{60}=500$ ms, where the iIRM method provided s slightly better result.
%%As we can see, the proposed method provided better results than the benchmark algorithms, except the only case where the iIRM method provided a slightly better SDR value for the time-varying RIR regarding Room 2 at $RT_{60}=500$ ms.
%%The experiments using the simulated RIRs validate that the proposed method results in better de-reverberation performance even for both the unseen type of room environment and the time-varying RIR conditions.

The average results using the static and time-varying real-measured RIRs from the C4DM database, in case of an early reverberation RIR filter length of $Q_e=32$, are shown in Tables \ref{C4DM_static_RIR} and \ref{C4DM_tv_RIR}, respectively.
As we can see, the proposed method provided the best results for all room types as well as for static and time-varying RIRs.
Interestingly, the dIRM method provided better results than the DSM and iIRM methods in general especially in terms of the SDR value, in contrast to the results found when using the simulated RIRs in Tables \ref{Habets_static_RIR} and \ref{Habets_tv_RIR}.
%The experiments using the real-measured RIRs validate that the proposed method not only well functions but also improves the de-reverberation performance in a more realistic acoustic scenario.

In the following, we comment on some additional experimental results, which we did not report in this paper due to space limitation.
First, we observed that the proposed method provided better performance than the benchmark algorithms for $Q_e=64$ and $128$, following similar trend to those reported above for $Q_e=32$.
Second, we were able to verify that estimating the early reverberant signal $y_n^E$ resulted in better de-reverberation performance than attempting to estimate the clean speech signal $s_n$.

%\setlength\extrarowheight{2pt}
%%\setlength\extrarowheight{1.5pt}
%\begin{table}[t!]
%%\footnotesize
%\scriptsize
%%\tiny
%%\fontsize{7}{8}\selectfont % same as scriptsize
%%\fontsize{6}{8}\selectfont
%%\fontsize{6.5}{7.5}\selectfont
%\caption{Average results for the static real-measured RIRs}
%\centering
%	%\begin{tabular}{c|c|c|c|c|c|c||c}
%	%\begin{tabular}{G|E|F|F|F|F|F||F}
%	%\begin{tabular}{E|F|F|F|F|F||F}
%	\begin{tabular}{c|H|H|H|H|H||H}
%		\hline
%		Room & \multirow{2}{*}{Eval.} & \multirow{2}{*}{Rev.} & \multirow{2}{*}{DSM} & \multirow{2}{*}{iIRM} & \multirow{2}{*}{dIRM} & \multirow{2}{*}{iFilt} \\ %\cline{7-8}
%		type & & & & & & \\
%		\hline \hline
%		\multirow{3}{*}{GreatHall}
%            	& SDR & -2.36 & -0.59 & 0.27 & 1.25 & \textbf{1.73} \\ %\cline{2-10}
%            	& ESTOI & 0.28 & 0.39 & 0.30 & 0.42 & \textbf{0.45} \\ %\cline{2-10}
%            	& SRMR & 1.38 & 2.36 & 2.14 & 2.50 & \textbf{3.17} \\ %\cline{2-8}
%		\hline %\hline
%		\multirow{3}{*}{Octagon}
%            	& SDR & -2.95 & -0.84 & 0.64 & 1.02 & \textbf{1.58} \\ %\cline{2-10}
%            	& ESTOI & 0.29 & 0.39 & 0.30 & 0.42 & \textbf{0.45} \\ %\cline{2-10}
%            	& SRMR & 1.28 & 2.19 & 2.21 & 2.26 & \textbf{2.89} \\ %\cline{2-8}
%		\hline %\hline
%		\multirow{3}{*}{Classroom}
%            	& SDR & -4.08 & -1.93 & -1.96 & -1.52 & \textbf{-0.19} \\ %\cline{2-10}
%            	& ESTOI & 0.19 & 0.36 & 0.23 & 0.36 & \textbf{0.40} \\ %\cline{2-10}
%            	& SRMR & 1.16 & 2.12 & 1.81 & 2.00 & \textbf{2.67} \\ %\cline{2-10}
%		\hline %\hline		
%	\end{tabular}
%	\label{C4DM_static_RIR}
%\end{table}

\setlength\extrarowheight{2pt}
\begin{table}[t!]
%\footnotesize
\scriptsize
%\tiny
%\fontsize{7}{8}\selectfont % same as scriptsize
%\fontsize{6}{8}\selectfont
%\fontsize{6.5}{7.5}\selectfont
\caption{Average results for the static real-measured RIRs}
\centering
	%\begin{tabular}{c|c|c|c|c|c|c||c}
	%\begin{tabular}{G|E|F|F|F|F|F||F}
	%\begin{tabular}{E|F|F|F|F|F||F}
	\begin{tabular}{c|H|H|H|H|H||H}
		\hline
		Room & \multirow{2}{*}{Eval.} & \multirow{2}{*}{Rev.} & \multirow{2}{*}{DSM} & \multirow{2}{*}{iIRM} & \multirow{2}{*}{dIRM} & \multirow{2}{*}{iFilt} \\ %\cline{7-8}
		type & & & & & & \\
		\hline \hline
		\multirow{3}{*}{GreatHall}
            	& SDR & -2.36 & -0.59 & 0.27 & 1.25 & \textbf{1.84} \\ %\cline{2-10}
            	& ESTOI & 0.28 & 0.39 & 0.30 & 0.42 & \textbf{0.45} \\ %\cline{2-10}
            	& SRMR & 1.38 & 2.36 & 2.14 & 2.50 & \textbf{3.32} \\ %\cline{2-8}
		\hline %\hline
		\multirow{3}{*}{Octagon}
            	& SDR & -2.95 & -0.84 & 0.64 & 1.02 & \textbf{1.62} \\ %\cline{2-10}
            	& ESTOI & 0.29 & 0.39 & 0.30 & 0.42 & \textbf{0.46} \\ %\cline{2-10}
            	& SRMR & 1.28 & 2.19 & 2.21 & 2.26 & \textbf{3.06} \\ %\cline{2-8}
		\hline %\hline
		\multirow{3}{*}{Classroom}
            	& SDR & -4.08 & -1.93 & -1.96 & -1.52 & \textbf{-0.16} \\ %\cline{2-10}
            	& ESTOI & 0.19 & 0.36 & 0.23 & 0.36 & \textbf{0.40} \\ %\cline{2-10}
            	& SRMR & 1.16 & 2.12 & 1.81 & 2.00 & \textbf{2.86} \\ %\cline{2-10}
		\hline %\hline		
	\end{tabular}
	\label{C4DM_static_RIR}
\end{table}

\section{Conclusion and future works}
We introduced a spectral-domain inverse filtering approach for single-channel speech de-reverberation using a DNN. 
The main goal was to better handle realistic reverberant conditions where the RIR filter is longer than the STFT analysis window.
To this end, we considered the CTF model for the reverberant speech signal.
In the proposed framework, we aimed at estimating the magnitude spectral coefficients of the early reverberant speech signal via inverse filtering of the CTF model.
The inverse filter was estimated based on the online U-net architecture, which consists of a fully-convolutional CAE with skip-connections.
We conducted experiments using both the simulated and real-measured RIRs.
Experiments showed that the proposed method provides better de-reverberation performance than the prevalent benchmark algorithms under various reverberation conditions, i.e., different levels of reverberation time, unseen type of room environment, static and time-varying RIR conditions, and for the simulated and real-measured RIRs.

Several avenues remain opened for future research.
First, we can extend the proposed method to a complex-valued U-net architecture to handle the phase components \cite{Choi19}.
Second, we can incorporate additional information, such as reverberation time \cite{Wu17} or late reverberation PSD \cite{Qi19}, into the proposed inverse filtering framework. 
Besides the above avenues, which are mainly for further improving the speech quality of the reverberant speech, it would also be of interest to evaluate experimentally the effects of the proposed de-reverberation algorithm when used as a front-end for automatic speech recognition.

\setlength\extrarowheight{2pt}
\begin{table}[t!]
%\footnotesize
\scriptsize
%\tiny
%\fontsize{7}{8}\selectfont % same as scriptsize
%\fontsize{6}{8}\selectfont
%\fontsize{6.5}{7.5}\selectfont
\caption{Average results for the time-varying real-measured RIRs}
%\caption{Average results for the time-varying measured RIR conditions}
\centering
	%\begin{tabular}{c|c|c|c|c|c|c||c}
	%\begin{tabular}{G|E|F|F|F|F|F||F}
	%\begin{tabular}{c|F|F|F|F|F||F}
	\begin{tabular}{c|H|H|H|H|H||H}
		\hline
		Room & \multirow{2}{*}{Eval.} & \multirow{2}{*}{Rev.} & \multirow{2}{*}{DSM} & \multirow{2}{*}{iIRM} & \multirow{2}{*}{dIRM} & \multirow{2}{*}{iFilt} \\ %\cline{7-8}
		type & & & & & & \\
		\hline \hline
		\multirow{3}{*}{GreatHall}
            	& SDR & -5.42 & -3.02 & -2.06 & -1.57 & \textbf{-1.37} \\ %\cline{2-10}
            	& ESTOI & 0.23 & 0.36 & 0.26 & 0.38 & \textbf{0.42} \\ %\cline{2-10}
            	& SRMR & 1.37 & 2.41 & 2.12 & 2.41 & \textbf{3.33} \\ %\cline{2-8}
		\hline %\hline
		\multirow{3}{*}{Octagon}
            	& SDR & -4.59 & -2.63 & -1.81 & -0.70 & \textbf{-0.45} \\ %\cline{2-10}
            	& ESTOI & 0.26 & 0.37 & 0.28 & 0.40 & \textbf{0.44} \\ %\cline{2-10}
            	& SRMR & 1.14 & 2.05 & 2.00 & 2.20 & \textbf{2.91} \\ %\cline{2-8}
		\hline %\hline
		\multirow{3}{*}{Classroom}
            	& SDR & -6.07 & -3.98 & -4.19 & -3.87 & \textbf{-2.54} \\ %\cline{2-10}
            	& ESTOI & 0.19 & 0.36 & 0.22 & 0.36 & \textbf{0.40} \\ %\cline{2-10}
            	& SRMR & 1.14 & 2.10 & 1.39 & 1.99 & \textbf{2.84} \\ %\cline{2-10}
		\hline %\hline		
	\end{tabular}
	\label{C4DM_tv_RIR}
\end{table}

% Below is an example of how to insert images. Delete the ``\vspace'' line,
% uncomment the preceding line ``\centerline...'' and replace ``imageX.ps''
% with a suitable PostScript file name.
% -------------------------------------------------------------------------
%\begin{figure}[htb]
%
%\begin{minipage}[b]{1.0\linewidth}
%  \centering
%  \centerline{\includegraphics[width=8.5cm]{image1}}
%%  \vspace{2.0cm}
%  \centerline{(a) Result 1}\medskip
%\end{minipage}
%%
%\begin{minipage}[b]{.48\linewidth}
%  \centering
%  \centerline{\includegraphics[width=4.0cm]{image3}}
%%  \vspace{1.5cm}
%  \centerline{(b) Results 3}\medskip
%\end{minipage}
%\hfill
%\begin{minipage}[b]{0.48\linewidth}
%  \centering
%  \centerline{\includegraphics[width=4.0cm]{image4}}
%%  \vspace{1.5cm}
%  \centerline{(c) Result 4}\medskip
%\end{minipage}
%%
%\caption{Example of placing a figure with experimental results.}
%\label{fig:res}
%%
%\end{figure}

% To start a new column (but not a new page) and help balance the last-page
% column length use \vfill\pagebreak.
% -------------------------------------------------------------------------

%\vfill
%\pagebreak

%\section{REFERENCES}
%\label{sec:ref}
%
% -------------------------------------------------------------------------
%\bibliographystyle{IEEEbib}
%\bibliography{strings,refs}

\end{document}